\documentclass[useAMS,usenatbib]{mn2e}

\usepackage{natbib}
\usepackage{graphicx}
\usepackage{multicol}
\usepackage{longtable}
\usepackage{amssymb}
\usepackage[width=1.0\textwidth]{caption}
\usepackage[usenames,dvipsnames]{color}
\RequirePackage[pdftex, plainpages = false, pdfpagelabels]{hyperref}
\RequirePackage{subfigure} %
\definecolor{Red}{rgb}{1.0,0,0}

\newcommand{\etal}    {{\it et al}}                          
\newcommand{\SLPJ}[4]{$^{#1}$#2$^{\rm{#3}}_{_{#4}}$}       
\newcommand{\nlo} [3]{#1#2$^{#3}$}                 

\newcommand{\Ii}      {~{\sc i}}
\newcommand{\II}      {~{\sc ii}}
\newcommand{\III}     {~{\sc iii}}

\newcommand{\otp}     {O$^{2+}$}
\newcommand{\Dosilon} {\rotatebox[origin=c]{180}{$\Upsilon$}}

\title[$\kappa$-distributed effective collision strengths for {\rm O}\III] {Effective collision
  strengths for excitation and de-excitation of nebular [O\III] optical and infrared lines with
  $\kappa$ distributed electron energies}
\author[P.J. Storey \& Taha Sochi]
{P.J. Storey$^{1}$, Taha Sochi$^{1}$\thanks{E-mail: t.sochi@ucl.ac.uk} \\
$^{1}$Department of Physics and Astronomy, University College London, Gower Street, London WC1E
6BT, UK}

\begin{document}

\date{Accepted  2015 March 3. Received 2015 February 14; in original form 2015 January 15}

\maketitle

\label{firstpage}

\begin{abstract}
We present effective collision strengths for electron excitation and de-excitation of the ten
forbidden transitions between the five lowest energy levels of the astronomically abundant
doubly-ionised oxygen ion, O$^{2+}$. The raw collision strength data were obtained from an R-matrix
intermediate coupling calculation using the Breit-Pauli relativistic approximation published
previously by the authors. The effective collision strengths were calculated with
$\kappa$-distributed electron energies and are tabulated as a function of the electron temperature
and $\kappa$.
\end{abstract}

\begin{keywords}
atomic data -- atomic processes -- radiation mechanisms: non-thermal -- planetary nebulae: general
-- infrared: general.
\end{keywords}

\section{Introduction} \label{Introduction}

The spectral lines of doubly-ionised oxygen, \otp, are important diagnostic tools in a variety of
astronomical and astrophysical situations, such as solar and nebular studies, due to the abundance
of this ion in these environments and the brightness of a number of its lines
\citep{AggarwalK1999}. They are used for instance to determine the oxygen abundance and other
physical conditions in the Milky Way and other galaxies out to substantial cosmological distances
\citep{Maiolino2008}. There has been a recent advocation \citep{NichollsDS2012,NichollsDSKP2013} of
the use of non Maxwell-Boltzmann (MB) electron energy distributions in the analysis of the spectra
of planetary nebulae (PNe) and H\II\ regions. It has been proposed that the discrepancy between the
results for elemental abundance and electron temperature obtained from the optical recombination
lines (ORLs) and those obtained from the collisionally-excited lines (CELs) could be resolved if
the free electron energies are described by a $\kappa$-distribution \citep{Vasyliunas1968} rather
than a MB one. \citet{StoreySSS22013} attempted to test this hypothesis using a subset of
dielectronic recombination lines to directly sample the free electron energy distribution. No
evidence was found for departures from a MB distribution in a small sample of PNe but the
uncertainties in the observational data made it impossible to rule out such departures.
\citet{ZhangLZ2014} modelled the shape of the Balmer continuum in the spectra of a sample of PNe
and concluded that both a model comprising two MB distributions of different temperatures or
$\kappa$-distributions matched the spectra within the observational errors. However,
\citet{StoreySSS42014} carried out the same modelling on the extreme PN, Hf 2-2, and concluded that
the probability that the spectrum corresponds to a $\kappa$-distribution is extremely low.

The present paper is part of a series of papers by the authors intended to make possible a
spectroscopic investigation of whether a $\kappa$ electron energy distribution can provide a
consistent explanation of the spectra of the thin and relatively cold plasma found in planetary
nebulae and H\II\ regions. The [O\III] lines are very strong in all PNe and the corresponding
recombination lines of O\II\ have also been recorded in many nebulae exhibiting varying degrees of
disagreement between ORL and CEL abundances and temperatures.  Indeed, the abundances derived from
the [O\III] forbidden lines and from the O\II\ recombination lines are one of the primary sources
for the ORL/CEL discrepancy. We therefore aim to provide in this series of papers
$\kappa$-dependent recombination coefficients for H\Ii\ and O\II\ and $\kappa$-dependent collision
strengths for the electron excitation and de-excitation of the [O\III] lines. With these
theoretical data we aim to answer the question whether any single $\kappa$-distribution can explain
the relative intensities of the [O\III] forbidden lines and corresponding O\II\ recombination lines
in a range of PNe and, if so, what is the resulting O$^{2+}$/H abundance ratio.

In \citet{StoreySB2014} the available theoretical parameters for electron collisional excitation of
the [O\III] forbidden lines were investigated and new collision strength data for the ten
transitions between the five lowest levels of the ground configuration were generated using a
Breit-Pauli relativistic model. In \citet{StoreySSS72015} the $\kappa$-distributed hydrogen
recombination data were provided. The purpose of the present paper is to investigate
$\kappa$-distributed effective collision strengths of the [O\III] forbidden transitions. What
remains is the O\II\ recombination with a $\kappa$ electron distribution which will be the subject
of a forthcoming paper.

\section{Collision Strengths}

There are a considerable number of studies dealing with the collision strengths of \otp\ and their
thermally-averaged values (e.g. \citet{BalujaBK1980, Aggarwal1983, Aggarwal1985, Aggarwal1993,
LennonB1994, AggarwalK1999, PalayNPE2012, StoreySB2014, MendozaB2014}). However, most of the past
studies provided a limited amount of effective collision strength data; moreover, except the last
one, the provided data are based on a Maxwell-Boltzmann electron energy distribution.
\citet{StoreySB2014} reviewed the existing theoretical work and concluded that there is good
agreement between their new collision strength results and all the past high-quality calculations.
However, they found significant differences with the work of \citet{PalayNPE2012} which, they
argued, were probably due to the small number of target states used by those authors, omitting key
terms of the 2p$^4$ electron configuration. The most recent work of \citet{MendozaB2014} also uses
a limited number of target states which omits the 2p$^4$ terms and the calculations were made with
an implementation of the Intermediate Coupling Frame Transformation (ICFT) method of
\citet{griffinetal98} which is potentially problematic for the deeply closed channels that are
encountered in low energy scattering from \otp\ \citep{StoreySB2014}. Recently,
\citet{AggarwalK2015} have suggested that the Be-sequence ICFT calculations of \citet{FMDZB2014}
might be unreliable due to their use of the ICFT method, quoting \citet{StoreySB2014} as evidence
of potential problems. It is important to emphasise that the problems encountered in the ICFT
calculations for O$^{2+}+e^{-}$ scattering were due to a specific combination of circumstances and
cannot be assumed to have occurred for any previous ICFT work. In the ICFT approach, closed channel
wavefunctions are integrated inwards to the boundary of the R-matrix inner region. The target
orbitals in the O$^{2+}+e^{-}$ calculation of \citet{StoreySB2014} were very compact, comprising
only $n=2$ spectroscopic orbitals and short range $n=3$ and $4$ correlation orbitals. For low
charge ions, deeply closed channels with small effective quantum number can arise and for
O$^{2+}+e^{-}$ the inward integration of these deeply closed channels to small values of radius
developed significant exponentially growing terms which distorted the results.

\begin{figure}
\centering{}
\includegraphics[height=7cm, width=8.5cm]{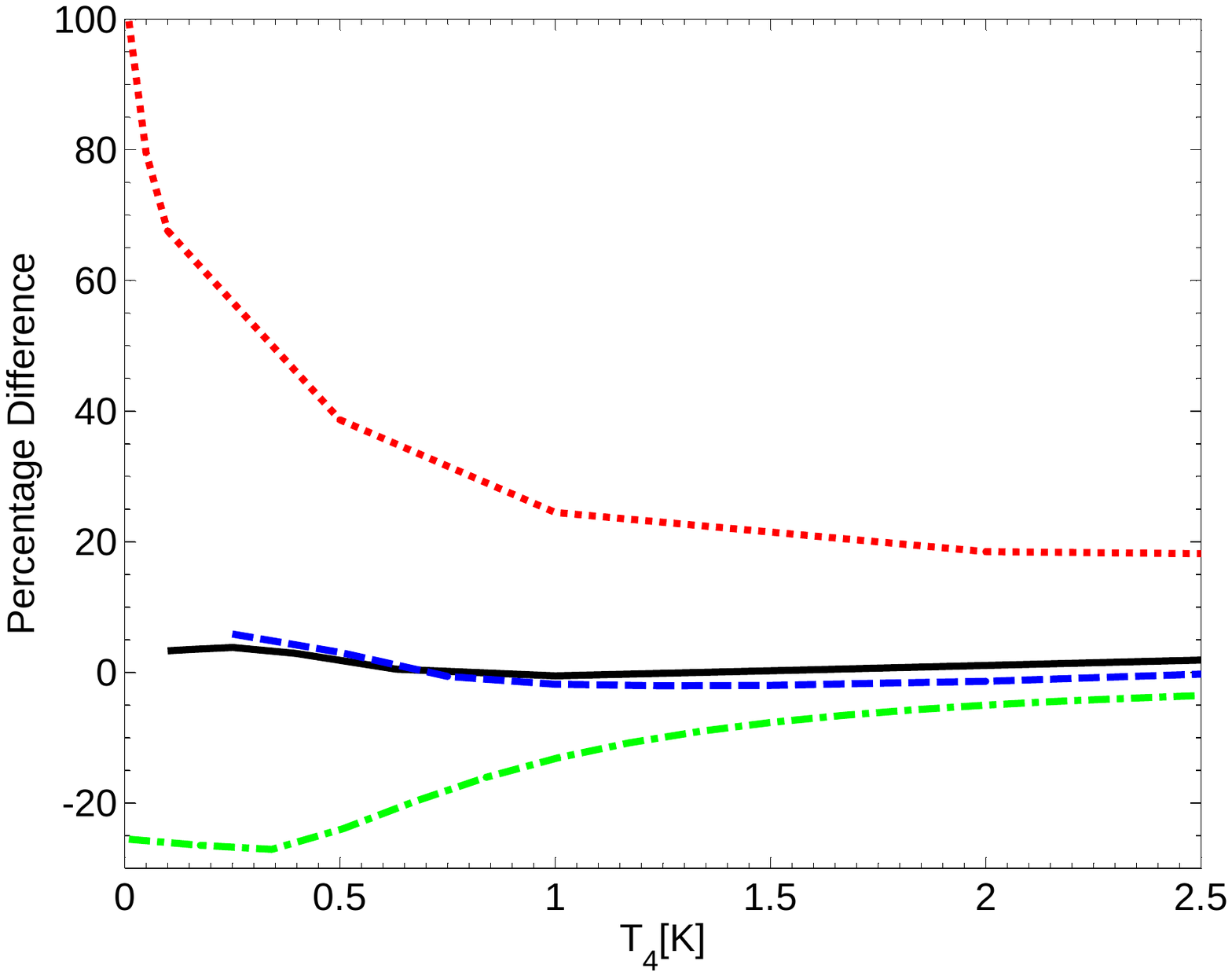}
\caption{Percentage difference of thermally averaged collision strengths from our 72-term
Breit-Pauli calculation versus temperature in 10$^4$~K for the transition 1-5. Results are from
\citet{LennonB1994} (solid black line), \citet{AggarwalK1999} (dashed blue line),
\citet{PalayNPE2012} (dotted red line), and \citet{MendozaB2014} (dash-dotted green line).}
\label{PlotMeBUpsilon152}
\end{figure}

In Figure~\ref{PlotMeBUpsilon152} we compare the MB-averaged collision strengths for the $^3$P$_0$
-- $^1$S$_0$ transition from \citet{LennonB1994}, \citet{AggarwalK1999}, \citet{PalayNPE2012} and
\citet{MendozaB2014} with those of \citet{StoreySB2014} as a function of electron temperature. As
in \citet{StoreySB2014} we show the percentage difference from the \citet{StoreySB2014} results in
each case. The values plotted for \citet{MendozaB2014} were calculated by us from their collision
strength data as privately supplied to us.  As with the \citet{PalayNPE2012} results, the
\citet{MendozaB2014} values differ significantly from those of \citet{StoreySB2014} and previous
large scale calculations. A possible source of the difference between the results of
\citet{StoreySB2014} and \citet{MendozaB2014} is that the latter's collision strengths extend to a
higher energy than the former's. However, on truncating the energy range of the
\citet{MendozaB2014} collision strengths to the same range used by \citet{StoreySB2014} we found
that the effective collision strengths are only marginally different in the temperature range
100--20000~K. We therefore continue to maintain the view expressed by \citet{StoreySB2014} that
there is no reason to believe that the consensus of the results of \citet{LennonB1994},
\citet{AggarwalK1999} and \citet{StoreySB2014}, which all agree within 10\%, is in error. A similar
pattern is seen in the other forbidden transitions among the lowest five levels.

\section{Theoretical Background}

The electron impact collision strength, $\Omega$, is a dimensionless quantity used to quantify the
intrinsic probability of collisional excitation and de-excitation in an atomic transition at a
particular electron energy. The effective collision strength is defined as the collision strength
averaged with respect to an electron energy distribution function, and hence it is obtained by
convolving the collision strength with a distribution function. The effective collision strength of
electron excitation, $\Upsilon$, from a lower state $i$ to an upper state $j$ is defined by

\begin{equation}
\Upsilon_{i\rightarrow j}=\frac{\sqrt{\pi}}{2}e^{\left(\frac{\Delta
E_{ij}}{k_{_B}T}\right)}\int_{0}^{\infty}\Omega_{ij}(\epsilon_{i})\,\,\sqrt{\frac{k_{_B}T}{\epsilon_{i}}}\,\,
f(\epsilon_{i})\,\, d\epsilon_{j}\end{equation}
where $T$ is the effective temperature, $k_{_B}$ is the Boltzmann constant, $\epsilon_{i}$ and
$\epsilon_{j}$ are the free electron energy relative to the states $i$ and $j$ respectively,
$\Delta E_{ij}$ ($=\epsilon_{i}-\epsilon_{j}$) is the energy difference between the two atomic
states, $\Omega_{ij}(\epsilon_{i})$ is the collision strength of the transition between the $i$ and
$j$ states, and $f(\epsilon_{i})$ is the electron energy distribution function. Similarly, the
effective collision strength of electron de-excitation, $\Dosilon$, from an upper state $j$ to a
lower state $i$ of a species is given by

\begin{equation}
\Dosilon_{j\rightarrow i}=\frac{\sqrt{\pi}}{2}\int_{0}^{\infty}\Omega_{ij}(\epsilon_{j})
\left(\frac{k_{_B}T}{\epsilon_{j}}\right)^{1/2}f(\epsilon_{j})\,\,d\epsilon_{j}
\end{equation}

The electron distribution function is usually assumed a Maxwell-Boltzmann which is given by

\begin{equation}
f_{_{{\rm MB}}}(\epsilon,T)=\frac{2\sqrt{\epsilon}}{\sqrt{\pi\left(k_{_B}T\right)^{3}}}\,\,
e^{-\frac{\epsilon}{k_{_B}T}}
\end{equation}
based on a thermodynamic equilibrium state. Similar thermal-averaging procedures with respect to
other electron distribution functions can also apply to reflect non-equilibrium states. The most
widely used non-equilibrium distribution function is the $\kappa$ distribution which is given, in
one of its common forms, by \citep{Vasyliunas1968, SummersT91}

\begin{equation}
f(\epsilon,T,\kappa)=\frac{2\sqrt{\epsilon}\,\,\,\Gamma(\kappa+1)}{\sqrt{\pi(k_{_B}T)^{3}(\kappa-\frac{3}{2})^{3}}\,\,\Gamma(\kappa-\frac{1}{2})\left(1+\frac{\epsilon}{(\kappa-\frac{3}{2})k_{_B}T}\right)^{\kappa+1}}
\end{equation}
where $\kappa$ is a parameter characterising the distribution and takes values in the interval
$(\frac{3}{2},\infty)$, and $\Gamma$ is the gamma function of the given arguments.

It should be remarked that $\Upsilon$ and $\Dosilon$ are identical, by definition, for the
Maxwell-Boltzmann distribution but are generally different for other types of electron distribution
function such as the $\kappa$ distribution. Also, the $\kappa$ distribution converges to the
Maxwell-Boltzmann distribution for large values of $\kappa$ although in many cases the convergence
practically occurs at moderate values of $\kappa$.

\section{Effective Collision Strengths} \label{Data}

We use the previously published raw collision strength data of \citet{StoreySB2014}. These were
generated using a 72-term atomic target with the Breit-Pauli Hamiltonian terms in the intermediate
coupling approximation as implemented in the UCL-Belfast-Strathclyde R-matrix code\footnote{{See
    Badnell: R-matrix write-up on WWW. URL: amdpp.phys.strath.ac.uk/tamoc/codes/serial/WRITEUP.}}
\citep{BerringtonEN1995}. The limits of the effective collision strength data are the same as the
limits of the original raw collision strength data that is: the data are related only to the ten
transitions between the five lowest levels of the ion and a free electron excitation energy up to
about 1.3~Rydberg. This provides effective collision strengths with a temperature up to 25000~K.
These limits were adopted mainly for the relevance of the data to thin and relatively cold plasma
found in planetary nebulae and H\II\ regions. \citet{MendozaB2014} have presented some graphical
comparisons of collision strengths averaged over $\kappa$-distributions for some [O\III] transitions
but very limited numerical results and only for MB distributions. Our data are therefore the first
comprehensive set of its kind in the public domain as far as we
know.

The data set accompanying this paper\footnote{The complete data generated in this work can be
obtained in electronic format with full precision from the Centre de Donn\'{e}es astronomiques de
Strasbourg (CDS) database.} consists of ten $\Upsilon$ files and ten $\Dosilon$ files where each
file is dedicated to one of the ten possible transitions between the five lowest levels of \otp\ as
given in Table~\ref{levelsTable}. In this regard we define a scaled Upsilon quantity, $\Upsilon_s$,
by
\begin{equation}\label{UpScaledEq}
\Upsilon_s = \Upsilon e^{\frac{-\Delta E_{ij}}{k_{_B}T}},
\end{equation}
to compensate for the large values of $\Upsilon$ when $T$ is low and $\Delta E_{ij}$ is large. We
also impose a cut-off limit for the acceptance of the scaled Upsilon values; the essence of this
condition is that any value that falls below $10^{-4}$ of the corresponding value at $T=10^4$~K for
the given $\kappa$ should be dropped. We use the indicator ``99999999'' to mark these dropped
entries for the purpose of keeping the rectangular structure of these tables. The physical basis
for imposing the cut-off condition is that since the scaled Upsilon is proportional to the
excitation rate coefficient, apart from a $T^{-1/2}$ factor, there is no practical use of the
excitation rate if it is very small compared to the generally-accepted standard thermal condition
in photoionised nebulae which is an MB distribution at $T=10^4$~K.

The collisional excitation rate coefficients, $q_{ij}$, can be obtained from the scaled Upsilons,
$\Upsilon_s$, using the following formula
\begin{equation}\label{aa}
 q_{ij}=\frac{2\sqrt{\pi} c \alpha a_0^2}{\omega_i}\sqrt{\frac{R}{k_{_B}T}} \Upsilon_s \simeq \frac{8.629\times10^{-6} \Upsilon_s}{\omega_i \sqrt{T}}, \ \ {\rm (cgs\,\,units)}
\end{equation}
where $\omega_i$ is the statistical weight of state $i$, $c$ is the speed of light in vacuum,
$\alpha$ is the fine structure constant, $a_0$ is the Bohr radius, and $R$ is the Rydberg energy
constant and where the temperature is in Kelvin. Similarly, the collisional de-excitation rate
coefficients, $q_{ji}$, can be obtained from the Downsilons, $\Dosilon$, using the following
formula
\begin{equation}\label{bb}
 q_{ji}=\frac{2\sqrt{\pi} c \alpha a_0^2}{\omega_j}\sqrt{\frac{R}{k_{_B}T}} \Dosilon \simeq \frac{8.629\times10^{-6} \Dosilon}{\omega_j \sqrt{T}} \ \ {\rm (cgs\,\,units)}
\end{equation}
where $\omega_j$ is the statistical weight of state $j$.

In each one of the aforementioned files, the 10-based logarithm of the effective collision
strengths, $\Upsilon_s$ and $\Dosilon$, are tabulated in a rectangular array where the columns
represent 10-based logarithmic electron temperatures [log$_{10}T$=2.0(0.025)4.3] while the rows
represent $\kappa$ which ranges between $1.6-10^{6}$ in unevenly-spaced periods [1.600(0.025)1.975,
2.0(0.1)2.9, 3.0(0.2)4.8, 5.0(0.5)9.5, 10(1)19, 20(2.5)47.5, 50(5)95, 100(25)175, 200(50)450,
500(100)900, 1000(1000)5000, 10000, 50000, 100000 and 1000000]. The $T$ and $\kappa$ grids are
constructed to satisfy the condition that the maximum percentage error in $\Upsilon_s$ or
$\Dosilon$ from linear interpolation as a function of $T$ or $\kappa$, between any two consecutive
log$_{10}$ values in the data files, should not exceed 1\% at the mid-point. More details about the
interpolation errors for these tables are given in Table \ref{errorTable}. The structure and
contents of the data files are fully explained in the ReadMe file that associates the data set. A
sample of these data for the 1-5 transition are presented in Table \ref{TableUp15} for the
logarithmic scaled Upsilon and in Table \ref{TableDo15} for the logarithmic Downsilon.

\begin{figure}
\centering %
\subfigure
{\begin{minipage}[b]{0.5\textwidth} \centering \includegraphics[height=7cm, width=8.5cm] {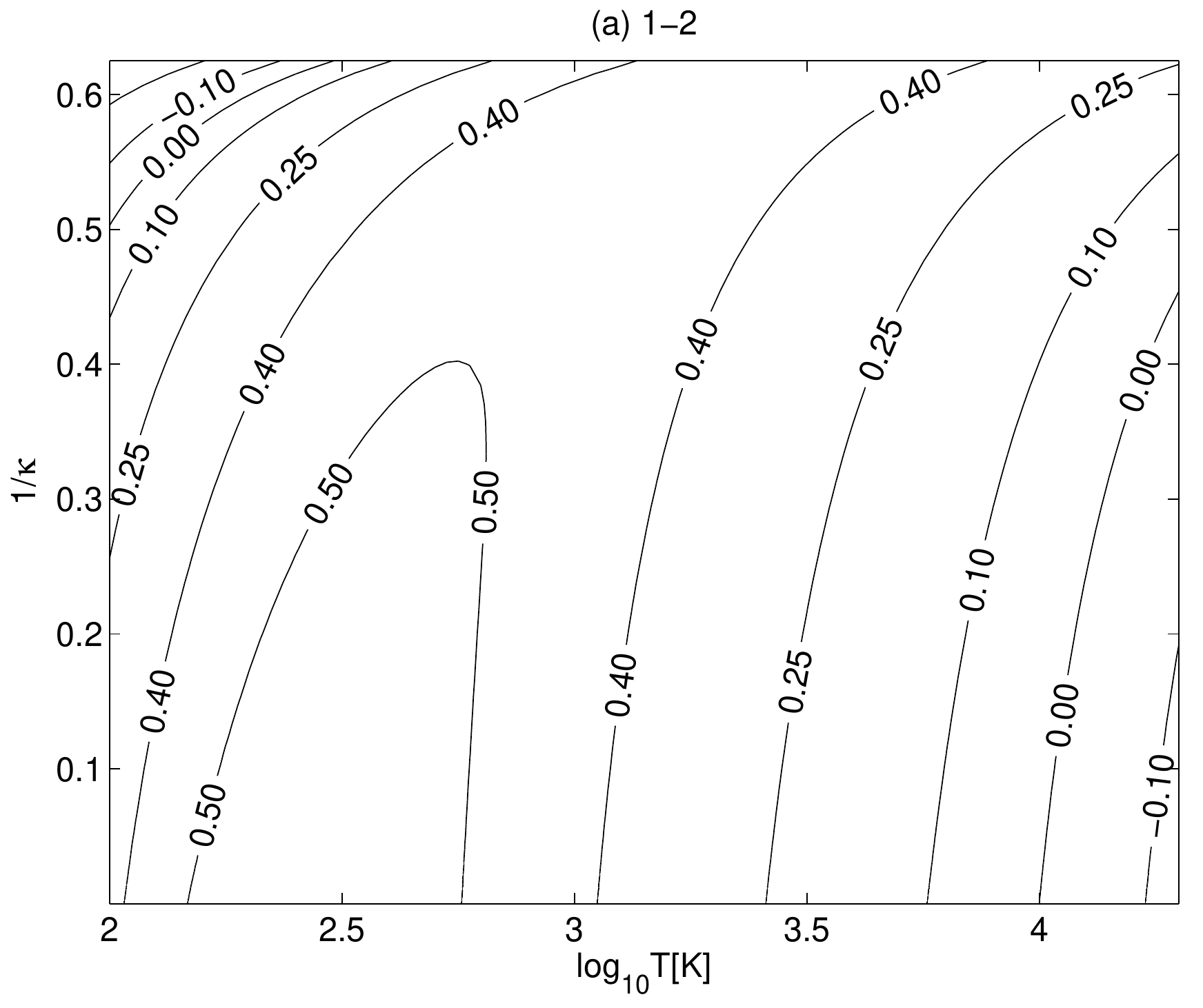}
\end{minipage}} \vspace{-0.3cm}
\centering %
\subfigure
{\begin{minipage}[b]{0.5\textwidth} \centering \includegraphics[height=7cm, width=8.5cm] {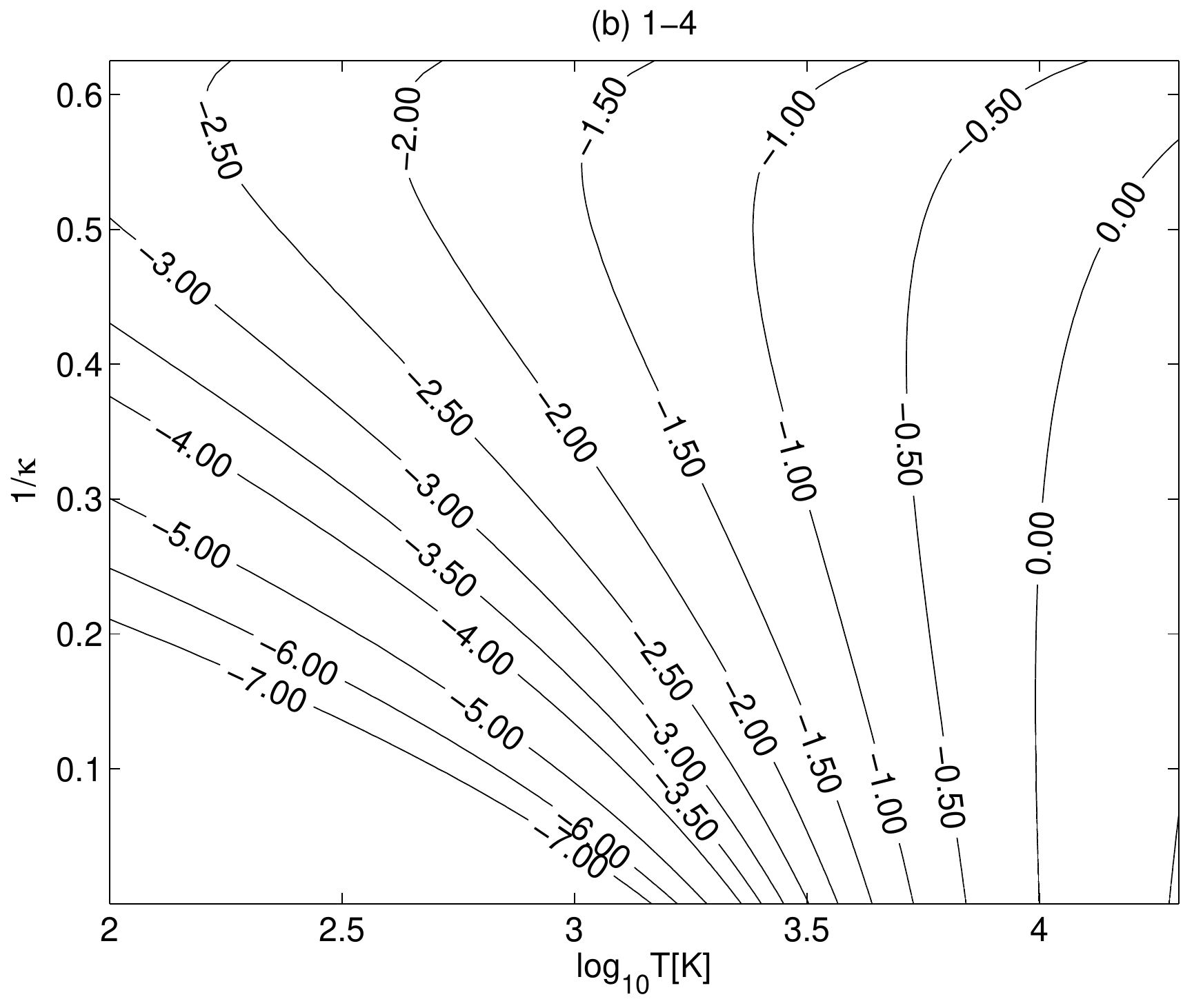}
\end{minipage}} \vspace{-0.3cm}
\centering %
\subfigure
{\begin{minipage}[b]{0.5\textwidth} \centering \includegraphics[height=7cm, width=8.5cm] {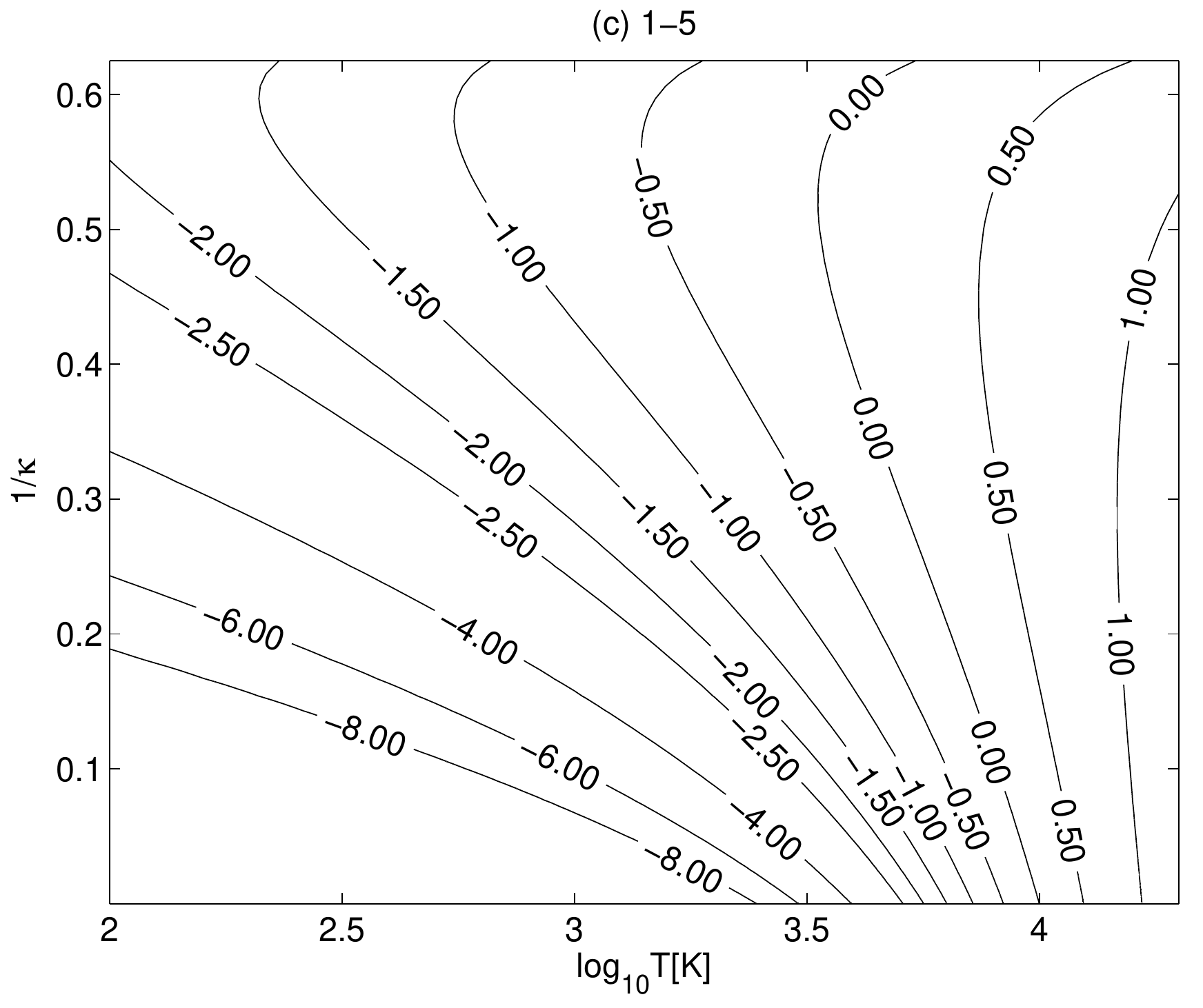}
\end{minipage}}
\caption{Contour plots for the log$_{10}$ scaled Upsilons (Equation \ref{UpScaledEq}) for the
transitions (a) 1-2, (b) 1-4 and (c) 1-5 as functions of log$_{10}T$ in K and $1/\kappa$. The
scaled Upsilon values were first normalised to the square root of their temperature ($\sqrt{T}$)
then to a reference value of Upsilon of a MB distribution at $T=10^4$~K. \label{ContourPlots}}
\end{figure}

In Figure~\ref{ContourPlots} we show contour plots of $\Upsilon_s/\sqrt{T}$ as a function of $T$
and $1/\kappa$ for three representative transitions. In this and subsequent contour plots the
values are base 10 logarithms of the relevant quantity, normalised to the value for an MB
distribution at $10^4$~K, which we will refer to as the ``standard PN conditions''. The plotted
quantity is proportional to the excitation rate coefficient for each transition. The 1-4 transition
is one of the means of exciting level 4, the $^1$D$_2$ level, which gives rise to the important
$\lambda\lambda 4959, 5007$ lines. The plot for this transition shows that the 1-4 excitation rate
coefficient at $10^4$~K is largely insensitive to the value of $\kappa$ for values of $\kappa$
larger than about 3. The picture is different for the 1-5 transition which excites the
$\lambda4363$ line which, by comparison with the $\lambda\lambda 4959, 5007$ lines is a temperature
diagnostic. The contour plot for this transition shows that the excitation rate coefficient is
sensitive to $\kappa$ in such a way that the standard PN value can occur at lower temperatures as
$\kappa$ decreases. Hence we can expect that the ratio of the intensities of the lines from level 4
and 5, which are commonly used as a temperature diagnostic, will also be sensitive to the value of
$\kappa$.

\begin{figure}
\centering{}
\includegraphics[height=7cm, width=8.5cm]{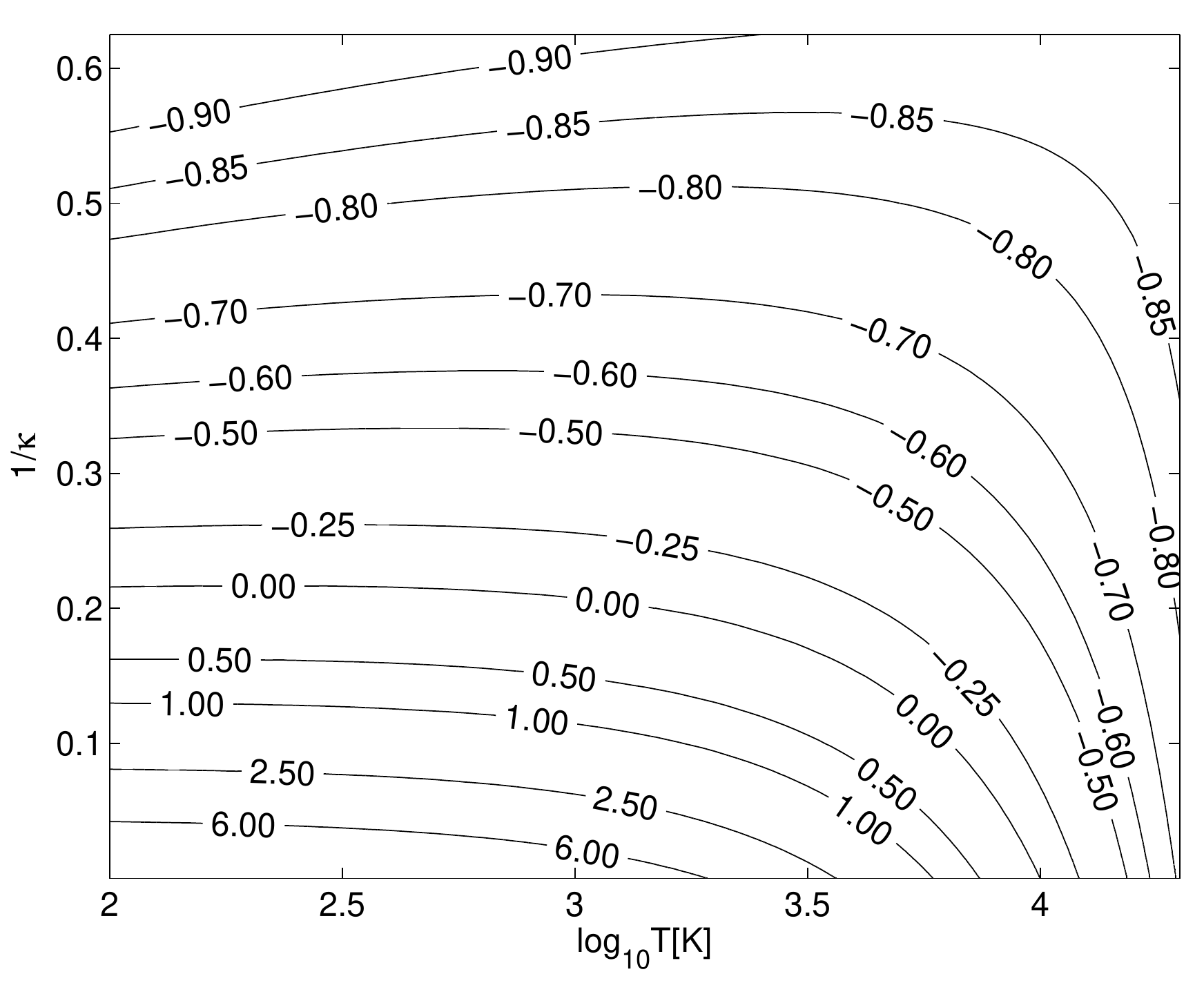}
\caption{A contour plot of log$_{10}[(\varepsilon(\lambda 4959)+\varepsilon(\lambda
5007))/\varepsilon(\lambda 4363)]$ as a function of log$_{10}T$ in K and $1/\kappa$, normalised to
the MB value at $T=10^4$~K.} \label{TempContour}
\end{figure}

In Figure~\ref{TempContour} we show a contour plot of the principal [O\III] temperature diagnostic,
the ratio of the sum of the emission coefficients, $\varepsilon$, of the two lines $\lambda\lambda
4959, 5007$ to the $\lambda 4363$ line at an electron number density of $10^4$~cm$^{-3}$. The
emission coefficients were computed using a model atom comprising the five energetically lowest
levels, and only electron collisional excitation and de-excitation and spontaneous radiative
transitions were included. The radiative transition probabilities were taken from
\citet{StoreyZ2000}, which we prefer to the more recent calculations of \citet{FischerT2004} for
O$^{2+}$ since the former calculation incorporates relativistic corrections to the magnetic dipole
operator.  Adding these corrections yields a more accurate value for the well observed $\lambda
4959$, $\lambda 5007$ branching ratio which was in disagreement with theory until the work of
\citet{StoreyZ2000}. The plot shows that the loci of the values of temperature and $\kappa$ where
the MB value is reproduced always lies at lower temperatures as $\kappa$ decreases and the
distribution moves away from MB. For example, the predicted ``standard'' value for the line ratio
of 194 is also obtained at $T\approx 6300$~K when $\kappa=10$. The predicted line intensities are
much lower at $6300$~K than at the standard temperature of $10^4$~K, so the number density of
O$^{2+}$ ions implied by a given measured line flux is larger.

\begin{figure}
\centering{}
\includegraphics[height=7cm, width=8.5cm]{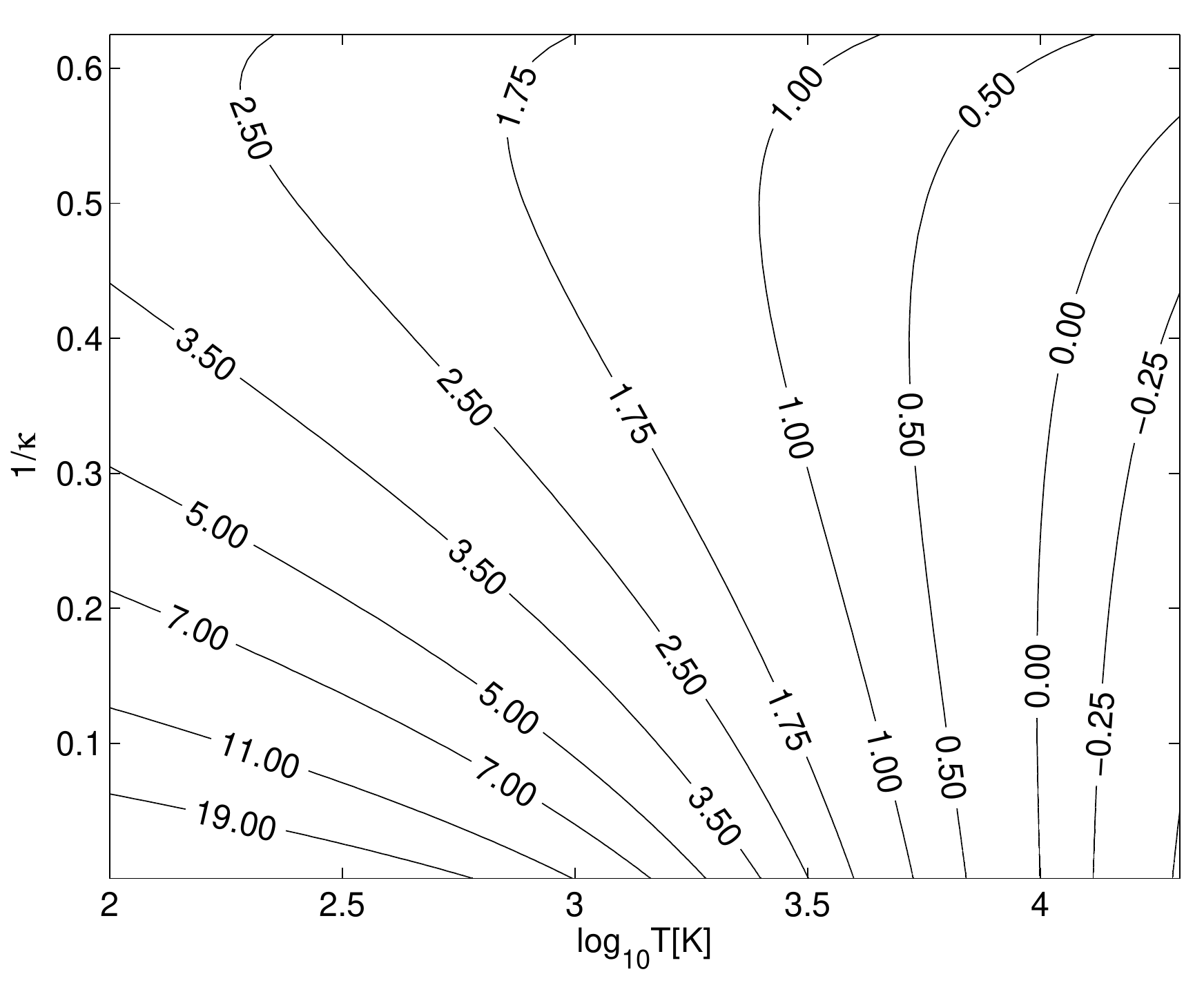}
\caption{A contour plot of the log$_{10}$ of the O$^{2+}$ number density derived from $\lambda
5007$ line as a function of log$_{10}T$ in K and $1/\kappa$, normalised to the value derived from
an MB distribution at $T=10^4$~K.} \label{AbundanceContour}
\end{figure}

In Figure~\ref{AbundanceContour} we show a contour plot of the O$^{2+}$ number density, assuming a
fixed intensity for the $\lambda5007$ line, as a function of temperature and $\kappa$. For the
example above, we find that at $T\approx 6300$~K and $\kappa=10$ the number density is 3.2 times
larger than that derived for standard conditions. The general conclusion therefore is that the
observed optical [O\III] line ratios can be modelled using a non-MB $\kappa$-distribution but with
lower temperature and higher derived O$^{2+}$ number density.

\begin{table} 
\caption{The five lowest energy levels of \otp\ with their experimental energies, $E_{\rm{ex}}$, in
wavenumbers (cm$^{-1}$). These values were obtained from the National Institute of Standards and
Technology database (www.nist.gov).} \label{levelsTable} \vspace{0.2cm} \centering
\begin{tabular}{|l|l|l|}
\hline
{\bf Index} & {\bf Level} & {\bf $E_{\rm{ex}}$} \\
\hline
   {\bf 1} & \nlo1s2 \nlo2s2 \nlo2p2 \SLPJ3P{}0 &       0.00 \\
   {\bf 2} & \nlo1s2 \nlo2s2 \nlo2p2 \SLPJ3P{}1 &     113.18 \\
   {\bf 3} & \nlo1s2 \nlo2s2 \nlo2p2 \SLPJ3P{}2 &     306.17 \\
   {\bf 4} & \nlo1s2 \nlo2s2 \nlo2p2 \SLPJ1D{}2 &   20273.27 \\
   {\bf 5} & \nlo1s2 \nlo2s2 \nlo2p2 \SLPJ1S{}0 &   43185.74 \\
\hline
\end{tabular}
\end{table}

\section{Conclusions} \label{Conclusions}

In the present paper, the effective collision strengths for electron excitation and de-excitation
between the lowest five levels of ${\rm O}^{2+}$ are computed with a $\kappa$ electron energy
distribution using a previously published set of collision strengths. Extensive tabulations of the
effective collision strengths as a function of temperature and $\kappa$ are provided. We also
illustrate and discuss the general behaviour of the most commonly used [O\III] visible lines as a
function of these two variables, and the effect on abundance determinations of adopting
non-Maxwellian distributions for the free electrons. This work is intended to provide the data
required for modelling and analysing optically thin plasma found mainly in planetary nebulae and
H\II\ regions, where it has been suggested that a non-MB $\kappa$ electron distribution may apply
and could provide a solution to the long standing problem in nebular physics of the contradiction
between the results of elemental abundance and electron temperature as obtained from optical
recombination lines versus those obtained from collisionally excited lines.

\section{Acknowledgment}

The work of PJS was supported in part by STFC (grant ST/J000892/1). The authors would like to thank
C. Mendoza and M.A. Bautista for sharing their data and atomic model.

\onecolumn \clearpage

\begin{table} 
\caption{Maximum and average error in interpolating the logarithmic values of scaled Upsilon and
Downsilon in $T$ and $\kappa$ where
MET=maximum absolute percentage error in $T$ interpolation,
AET=average absolute percentage error in $T$ interpolation,
MEK=maximum absolute percentage error in $\kappa$ interpolation,
and AEK=average absolute percentage error in $\kappa$ interpolation.} \label{errorTable}
\vspace{0.2cm} \centering
\begin{tabular}{cc|cccc|c|cccc}
\hline
    {\bf } &            &               \multicolumn{ 4}{|c}{{\bf Upsilon}} &            &             \multicolumn{ 4}{|c}{{\bf Downsilon}} \\
\hline
{\bf Transition} &            &  {\bf MET} &  {\bf AET} &  {\bf MEK} &  {\bf AEK} &            &  {\bf MET} &  {\bf AET} &  {\bf MEK} &  {\bf AEK} \\
\hline
  {\bf 1-2} &            &      0.069 &      0.012 &      0.562 &      0.021 &            &      0.012 &      0.002 &      0.367 &      0.024 \\

  {\bf 1-3} &            &      0.178 &      0.025 &      0.562 &      0.024 &            &      0.017 &      0.001 &      0.308 &      0.020 \\

  {\bf 1-4} &            &      0.464 &      0.074 &      0.563 &      0.072 &            &      0.017 &      0.002 &      0.319 &      0.021 \\

  {\bf 1-5} &            &      0.586 &      0.088 &      0.872 &      0.093 &            &      0.014 &      0.002 &      0.313 &      0.021 \\

  {\bf 2-3} &            &      0.112 &      0.018 &      0.562 &      0.022 &            &      0.017 &      0.001 &      0.312 &      0.021 \\

  {\bf 2-4} &            &      0.461 &      0.074 &      0.563 &      0.072 &            &      0.017 &      0.002 &      0.319 &      0.021 \\

  {\bf 2-5} &            &      0.584 &      0.088 &      0.868 &      0.093 &            &      0.014 &      0.002 &      0.313 &      0.021 \\

  {\bf 3-4} &            &      0.459 &      0.074 &      0.563 &      0.071 &            &      0.017 &      0.002 &      0.319 &      0.021 \\

  {\bf 3-5} &            &      0.581 &      0.088 &      0.860 &      0.093 &            &      0.014 &      0.002 &      0.313 &      0.021 \\

  {\bf 4-5} &            &      0.455 &      0.078 &      0.571 &      0.075 &            &      0.017 &      0.005 &      0.304 &      0.020 \\
\hline
\end{tabular}
\end{table}

\begin{table} 
\caption{Sample log$_{10}$ of scaled Upsilon data for the 1-5 transition.} \label{TableUp15}
\vspace{0.2cm} \centering
\begin{tabular}{r|rrrrrrrrrrrrr}
\hline
   {\bf $T$} & {\bf 2.000} & {\bf 2.025} & {\bf 2.050} & {\bf 2.075} & {\bf 2.100} & {\bf 2.125} & {\bf 2.150} & {\bf 2.175} & {\bf 2.200} & {\bf 2.225} & {\bf 2.250} & {\bf 2.275} & {\bf 2.300} \\
\hline
{\bf $\kappa$} &            &            &            &            &            &            &            &            &            &            &            &            &            \\

{\bf 1.600} &     -7.085 &     -7.045 &     -7.005 &     -6.965 &     -6.925 &     -6.885 &     -6.845 &     -6.805 &     -6.765 &     -6.725 &     -6.685 &     -6.645 &     -6.605 \\

{\bf 1.625} &     -7.063 &     -7.022 &     -6.981 &     -6.941 &     -6.900 &     -6.860 &     -6.819 &     -6.778 &     -6.738 &     -6.697 &     -6.657 &     -6.616 &     -6.575 \\

{\bf 1.650} &     -7.056 &     -7.015 &     -6.974 &     -6.933 &     -6.892 &     -6.850 &     -6.809 &     -6.768 &     -6.727 &     -6.685 &     -6.644 &     -6.603 &     -6.562 \\

{\bf 1.675} &     -7.061 &     -7.019 &     -6.977 &     -6.935 &     -6.894 &     -6.852 &     -6.810 &     -6.768 &     -6.726 &     -6.684 &     -6.642 &     -6.601 &     -6.559 \\

{\bf 1.700} &     -7.073 &     -7.031 &     -6.988 &     -6.946 &     -6.903 &     -6.861 &     -6.818 &     -6.776 &     -6.733 &     -6.691 &     -6.648 &     -6.606 &     -6.563 \\

{\bf 1.725} &     -7.091 &     -7.048 &     -7.005 &     -6.961 &     -6.918 &     -6.875 &     -6.832 &     -6.789 &     -6.746 &     -6.703 &     -6.660 &     -6.617 &     -6.574 \\

{\bf 1.750} &     -7.113 &     -7.069 &     -7.025 &     -6.982 &     -6.938 &     -6.894 &     -6.850 &     -6.807 &     -6.763 &     -6.719 &     -6.676 &     -6.632 &     -6.588 \\

{\bf 1.775} &     -7.138 &     -7.094 &     -7.050 &     -7.005 &     -6.961 &     -6.917 &     -6.872 &     -6.828 &     -6.784 &     -6.739 &     -6.695 &     -6.651 &     -6.606 \\

{\bf 1.800} &     -7.167 &     -7.122 &     -7.077 &     -7.032 &     -6.987 &     -6.942 &     -6.897 &     -6.852 &     -6.807 &     -6.762 &     -6.717 &     -6.672 &     -6.627 \\

{\bf 1.825} &     -7.197 &     -7.152 &     -7.106 &     -7.061 &     -7.015 &     -6.969 &     -6.924 &     -6.878 &     -6.833 &     -6.787 &     -6.741 &     -6.696 &     -6.650 \\

{\bf 1.850} &     -7.230 &     -7.184 &     -7.137 &     -7.091 &     -7.045 &     -6.999 &     -6.953 &     -6.906 &     -6.860 &     -6.814 &     -6.768 &     -6.722 &     -6.675 \\

{\bf 1.875} &     -7.264 &     -7.217 &     -7.170 &     -7.124 &     -7.077 &     -7.030 &     -6.983 &     -6.936 &     -6.889 &     -6.843 &     -6.796 &     -6.749 &     -6.702 \\

{\bf 1.900} &     -7.300 &     -7.252 &     -7.205 &     -7.157 &     -7.110 &     -7.062 &     -7.015 &     -6.967 &     -6.920 &     -6.872 &     -6.825 &     -6.778 &     -6.730 \\

{\bf 1.925} &     -7.336 &     -7.288 &     -7.240 &     -7.192 &     -7.144 &     -7.096 &     -7.048 &     -7.000 &     -6.952 &     -6.904 &     -6.855 &     -6.807 &     -6.759 \\

{\bf 1.950} &     -7.374 &     -7.325 &     -7.277 &     -7.228 &     -7.179 &     -7.130 &     -7.082 &     -7.033 &     -6.984 &     -6.936 &     -6.887 &     -6.838 &     -6.790 \\

{\bf 1.975} &     -7.412 &     -7.363 &     -7.314 &     -7.264 &     -7.215 &     -7.166 &     -7.117 &     -7.067 &     -7.018 &     -6.969 &     -6.919 &     -6.870 &     -6.821 \\

{\bf 2.000} &     -7.452 &     -7.402 &     -7.352 &     -7.302 &     -7.252 &     -7.202 &     -7.152 &     -7.102 &     -7.052 &     -7.002 &     -6.952 &     -6.902 &     -6.853 \\
\hline
\end{tabular}
\end{table}

\begin{table} 
\caption{Sample log$_{10}$ of Downsilon data for the 1-5 transition.} \label{TableDo15}
\vspace{0.2cm} \centering
\begin{tabular}{r|rrrrrrrrrrrrr}
\hline
   {\bf $T$} & {\bf 2.000} & {\bf 2.025} & {\bf 2.050} & {\bf 2.075} & {\bf 2.100} & {\bf 2.125} & {\bf 2.150} & {\bf 2.175} & {\bf 2.200} & {\bf 2.225} & {\bf 2.250} & {\bf 2.275} & {\bf 2.300} \\
\hline
{\bf $\kappa$} &            &            &            &            &            &            &            &            &            &            &            &            &            \\

{\bf 1.600} &     -1.123 &     -1.117 &     -1.111 &     -1.105 &     -1.100 &     -1.096 &     -1.091 &     -1.088 &     -1.084 &     -1.081 &     -1.078 &     -1.075 &     -1.073 \\

{\bf 1.625} &     -1.145 &     -1.140 &     -1.135 &     -1.131 &     -1.128 &     -1.124 &     -1.121 &     -1.119 &     -1.116 &     -1.114 &     -1.112 &     -1.110 &     -1.108 \\

{\bf 1.650} &     -1.166 &     -1.162 &     -1.158 &     -1.155 &     -1.153 &     -1.150 &     -1.148 &     -1.146 &     -1.144 &     -1.142 &     -1.141 &     -1.139 &     -1.138 \\

{\bf 1.675} &     -1.185 &     -1.182 &     -1.179 &     -1.177 &     -1.175 &     -1.173 &     -1.171 &     -1.169 &     -1.168 &     -1.167 &     -1.165 &     -1.164 &     -1.163 \\

{\bf 1.700} &     -1.203 &     -1.200 &     -1.198 &     -1.196 &     -1.194 &     -1.193 &     -1.191 &     -1.190 &     -1.189 &     -1.188 &     -1.187 &     -1.186 &     -1.185 \\

{\bf 1.725} &     -1.219 &     -1.217 &     -1.215 &     -1.213 &     -1.212 &     -1.210 &     -1.209 &     -1.208 &     -1.207 &     -1.206 &     -1.206 &     -1.205 &     -1.204 \\

{\bf 1.750} &     -1.233 &     -1.231 &     -1.230 &     -1.228 &     -1.227 &     -1.226 &     -1.225 &     -1.224 &     -1.224 &     -1.223 &     -1.222 &     -1.222 &     -1.221 \\

{\bf 1.775} &     -1.246 &     -1.245 &     -1.244 &     -1.242 &     -1.241 &     -1.240 &     -1.240 &     -1.239 &     -1.238 &     -1.238 &     -1.237 &     -1.237 &     -1.236 \\

{\bf 1.800} &     -1.258 &     -1.257 &     -1.256 &     -1.255 &     -1.254 &     -1.253 &     -1.252 &     -1.252 &     -1.251 &     -1.251 &     -1.250 &     -1.250 &     -1.250 \\

{\bf 1.825} &     -1.269 &     -1.268 &     -1.267 &     -1.266 &     -1.266 &     -1.265 &     -1.264 &     -1.264 &     -1.263 &     -1.263 &     -1.262 &     -1.262 &     -1.262 \\

{\bf 1.850} &     -1.279 &     -1.278 &     -1.278 &     -1.277 &     -1.276 &     -1.275 &     -1.275 &     -1.274 &     -1.274 &     -1.274 &     -1.273 &     -1.273 &     -1.273 \\

{\bf 1.875} &     -1.289 &     -1.288 &     -1.287 &     -1.286 &     -1.286 &     -1.285 &     -1.285 &     -1.284 &     -1.284 &     -1.284 &     -1.283 &     -1.283 &     -1.283 \\

{\bf 1.900} &     -1.297 &     -1.297 &     -1.296 &     -1.295 &     -1.295 &     -1.294 &     -1.294 &     -1.293 &     -1.293 &     -1.293 &     -1.292 &     -1.292 &     -1.292 \\

{\bf 1.925} &     -1.305 &     -1.305 &     -1.304 &     -1.303 &     -1.303 &     -1.302 &     -1.302 &     -1.302 &     -1.301 &     -1.301 &     -1.301 &     -1.301 &     -1.301 \\

{\bf 1.950} &     -1.313 &     -1.312 &     -1.312 &     -1.311 &     -1.311 &     -1.310 &     -1.310 &     -1.309 &     -1.309 &     -1.309 &     -1.309 &     -1.309 &     -1.308 \\

{\bf 1.975} &     -1.320 &     -1.319 &     -1.319 &     -1.318 &     -1.318 &     -1.317 &     -1.317 &     -1.317 &     -1.316 &     -1.316 &     -1.316 &     -1.316 &     -1.316 \\

{\bf 2.000} &     -1.326 &     -1.326 &     -1.325 &     -1.325 &     -1.324 &     -1.324 &     -1.324 &     -1.323 &     -1.323 &     -1.323 &     -1.323 &     -1.323 &     -1.322 \\
\hline
\end{tabular}
\end{table}


\begin{thebibliography}{99}

\bibitem[\protect\citeauthoryear{Aggarwal}{1983}]{Aggarwal1983}
Aggarwal K.M., 1983, ApJS, 52, 387

\bibitem[\protect\citeauthoryear{Aggarwal}{1985}]{Aggarwal1985}
Aggarwal K.M., 1985, A\&A, 146, 149

\bibitem[\protect\citeauthoryear{Aggarwal}{1993}]{Aggarwal1993}
Aggarwal K.M., 1993, ApJS, 85, 197

\bibitem[\protect\citeauthoryear{Aggarwal \& Keenan}{1999}]{AggarwalK1999}
Aggarwal K.M., Keenan F.P., 1999, ApJS, 123, 311

\bibitem[\protect\citeauthoryear{Aggarwal \& Keenan}{2015}]{AggarwalK2015}
Aggarwal K.M., Keenan F.P., 2015, arXiv:1501.00808

\bibitem[\protect\citeauthoryear{Baluja \etal}{1980}]{BalujaBK1980}
Baluja K.L., Burke P.G., Kingston A.E., 1980, J. Phys. B, 13, 829

\bibitem[\protect\citeauthoryear{Berrington \etal}{1995}]{BerringtonEN1995}
Berrington K.A., Eissner W.B., Norrington P.H., 1995, Comp. Phys. Comm., 92, 290

\bibitem[\protect\citeauthoryear{Fern\'{a}ndez-Menchero \etal}{2014}]{FMDZB2014}
Fern\'{a}ndez-Menchero L., Del Zanna G., Badnell N.R., 2014, A\&A, 566, A104

\bibitem[\protect\citeauthoryear{Fischer \& Tachiev}{2004}]{FischerT2004}
Fischer C.F., Tachiev G., 2004, At. Data. Nucl. Data Tables, 87, 1

\bibitem[\protect\citeauthoryear{Griffin, Badnell \& Pindzola}{1998}]{griffinetal98}
Griffin D.C., Badnell N.R., Pindzola M.S., 1998, J. Phys. B, 31, 3713

\bibitem[\protect\citeauthoryear{Lennon \& Burke}{1994}]{LennonB1994}
Lennon D.J., Burke V.M., 1994, A\&AS, 103, 273

\bibitem[\protect\citeauthoryear{Maiolino \etal}{2008}]{Maiolino2008}
Maiolino R., Nagao T., Grazian A., \etal, 2008, A\&A, 488, 463

\bibitem[\protect\citeauthoryear{Mendoza \& Bautista}{2014}]{MendozaB2014}
Mendoza C., Bautista M.A., 2014, ApJ, 785, 91

\bibitem[\protect\citeauthoryear{Nicholls \etal}{2012}]{NichollsDS2012}
Nicholls D.C., Dopita M.A., Sutherland R.S., 2012, ApJ, 752, 148

\bibitem[\protect\citeauthoryear{Nicholls \etal}{2013}]{NichollsDSKP2013}
Nicholls D.C., Dopita M.A., Sutherland R.S., Kewley L.J., Palay E., 2013, ApJS, 207, 1

\bibitem[\protect\citeauthoryear{Palay \etal}{2012}]{PalayNPE2012}
Palay E., Nahar S.N., Pradhan A.K., Eissner W., 2012, MNRAS Let., 423, L35

\bibitem[\protect\citeauthoryear{Storey \& Sochi}{2013}]{StoreySSS22013}
Storey P.J., Sochi T., 2013, MNRAS, 430, 599

\bibitem[\protect\citeauthoryear{Storey \& Sochi}{2014}]{StoreySSS42014}
Storey P.J., Sochi T., 2014, MNRAS, 440, 2581

\bibitem[\protect\citeauthoryear{Storey \etal}{2014}]{StoreySB2014}
Storey P.J., Sochi T., Badnell N.R., 2014, MNRAS,  441, 3028

\bibitem[\protect\citeauthoryear{Storey \& Sochi}{2015}]{StoreySSS72015}
Storey P.J., Sochi T., 2015, MNRAS, 446, 1864

\bibitem[\protect\citeauthoryear{Storey \& Zeippen}{2000}]{StoreyZ2000}
Storey P.J., Zeippen, C., 2000, MNRAS, 312, 813

\bibitem[\protect\citeauthoryear{Summers \& Thorne}{1991}]{SummersT91}
Summers D., Thorne R.M. 1991, Phys. Fluids B, 3, 1835

\bibitem[\protect\citeauthoryear{Vasyliunas}{1968}]{Vasyliunas1968}
Vasyliunas V.M., 1968, JGR, 73, 2839

\bibitem[\protect\citeauthoryear{Zhang \etal}{2014}]{ZhangLZ2014}
Zhang Y., Liu X-W., Zhang B., 2014, ApJ, 780, 93

\end{thebibliography}
\end{document}